\documentstyle[twocolumn,prl,aps,epsf]{revtex}
\def\beq{\begin{equation}}
\def\eeq{\end{equation}}
\def\beqn{\begin{eqnarray}}
\def\eeqn{\end{eqnarray}}
\def\bea{\begin{eqnarray}}
\def\eea{\end{eqnarray}} 
\def\be{\begin{equation}}
\def\ee{\end{equation}} 
\begin{document}                                                
\begin{titlepage}
\begin{flushright}FSU-HEP-2001-0601
\end{flushright} 
\begin{flushright}BNL-HET-01/20
\end{flushright}
\begin{flushright}hep-ph/0107101
\end{flushright}
\vspace{2truecm}
\begin{center}
{\large\bf
Next-to-Leading Order Results\\
\vskip .2in 
 for \boldmath$t {\overline t} h$\unboldmath
~Production at the Tevatron
}
\\
\vspace{1in}
{\bf L.~Reina}\\
{\it  Physics Department, Florida State University,\\ 
Tallahassee, FL 32306, USA}
\\ 
\vspace{.25in}
{\bf S.~Dawson}\\
{\it Physics Department, Brookhaven National Laboratory,\\
Upton, NY 11973, USA}
\vspace{1in}  
\end{center}
\begin{abstract} 
  We compute the ${\cal O}(\alpha_s^3)$ total cross section for the
  process $p {\overline p} \rightarrow t {\overline t} h$ in the
  Standard Model, at $\sqrt{s_H}\!=\!2$~TeV. The next-to-leading order
  corrections drastically reduce the renormalization and factorization
  scale dependence of the Born cross section and slightly decrease the
  total cross section for renormalization and factorization scales
  between $m_t$ and $2m_t$.  \clearpage
\end{abstract}
\end{titlepage} 

\title{Next-to-Leading Order Results 
 for \boldmath$t {\overline t} h$\unboldmath
~Production at the Tevatron}

\author{ L. Reina$^{a}$ and  S. Dawson$^{b}$}
\bigskip
\address{
a) Physics Department, Florida State University, 
Tallahassee, FL 32306, USA\\
b) Department of Physics, Brookhaven National Laboratory,
Upton, New York 11973-5000, USA}

\date{\today}
\maketitle
\begin{abstract} 
%************
  We compute the ${\cal O}(\alpha_s^3)$ total cross section for the
  process $p {\overline p} \rightarrow t {\overline t} h$ in the
  Standard Model, at $\sqrt{s_H}\!=\!2$~TeV. The next-to-leading order
  corrections drastically reduce the renormalization and factorization
  scale dependence of the Born cross section and slightly decrease
  the total cross section for renormalization and factorization scales
  between $m_t$ and $2m_t$.
%************
\end{abstract}
\begin{narrowtext}

%\section{Introduction}

  {\bf 1.} Among the most important goals of present and future colliders is
  the study of the electroweak symmetry breaking mechanism and the
  origin of fermion masses.  If the introduction of one or more Higgs
  fields is responsible for the breaking of the electroweak symmetry,
  then at least one Higgs boson should be relatively light, and
  certainly in the range of energies of present (Tevatron) or future
  (LHC) hadron colliders.  The present lower bounds on the Higgs mass
  have been set by LEP to be $M_h\!>\! 113.5$~GeV~\cite{leph} for the
  Standard Model (SM) Higgs boson ($h$), and $M_{h^0,
    A^0}\!>\!91$~GeV~\cite{lephwg} for the light scalar ($h^0$) and
  pseudoscalar ($A^0$) Higgs bosons of the minimal supersymmetric
  standard model (MSSM).  At the same time, precision fits to SM
  results indirectly point to the existence of a light Higgs boson,
  $M_h\!<\!212-236$~GeV~\cite{lepewwg}, while the MSSM requires the
  existence of a scalar Higgs boson lighter than about
  $130$~GeV~\cite{Heinemeyer:1999np}.  Therefore, the possibility of a
  Higgs boson discovery in the mass range around $115-130$~GeV seems
  increasingly likely.
  
  In this context, the Tevatron will play a crucial role and will have
  the opportunity to discover a Higgs boson in the mass range between
  the experimental lower bound and about 180~GeV~\cite{Carena:2000yx}.
  The dominant Higgs production modes at the Tevatron are gluon-gluon
  fusion, $gg\rightarrow h$, and the associated production with a weak
  boson, $q {\overline q}\rightarrow Wh,~Zh$ .  Because of small event
  rates and large backgrounds, the Higgs search in these channels can
  be problematic, requiring the highest possible luminosity.  It is
  therefore necessary to investigate all possible production channels,
  in the effort to fully exploit the range of opportunities offered by
  the available statistics.
  
  Recently, attention has been drawn to the possibility of detecting a
  Higgs signal in association with a pair of top-antitop quarks, i.e.
  in $p {\overline p} \rightarrow t {\overline t}
  h$~\cite{Goldstein:2001bp}. This production mode can play a role
  almost over the entire Higgs mass range accessible at the Tevatron.
  Although it has a small event rate, $\sim 1-5$~fb for a SM like
  Higgs, the signature ($W^+W^- b {\overline b} b{\overline b }$) is
  quite spectacular.  Furthermore, at the Tevatron (unlike at the
  LHC), the signal and background for this process have quite
  different shapes.  The statistics are too low to allow any direct
  measurement of the top Yukawa couplings, but recent studies
  \cite{incandela} indicate that this channel can reduce the
  luminosity required for a Higgs discovery at Run II of the Tevatron
  by as much as 15-20\%.
  
  Up to now the cross section for $p\bar p\rightarrow t\bar th$ has
  been known only at tree level. As for any other hadronic process,
  first order QCD corrections are expected to be important and are
  crucial in order to reduce the dependence of the cross section on
  the renormalization and factorization scales.  In this letter we
  present the results of our calculation of the NLO QCD corrections to
  the total cross section for $p {\overline p} \rightarrow t
  {\overline t} h$ in the Standard Model, at the Tevatron.  A detailed
  review of the calculation will be presented
  elsewhere~\cite{lr2}. We find good agreement with the analogous
  results presented in Ref.~\cite{been_etal}.
  \vspace{0.5truecm}

%\section{Basics}
  
{\bf 2.} The total cross section for $p {\overline p} \rightarrow t
  {\overline t} h$ at ${\cal O}(\alpha_s^3)$ can be written as:
\begin{eqnarray}
\sigma(p {\overline p} \rightarrow t {\overline t} h)_{NLO} 
&=&\sum_{ij}\int dx_1 dx_2 
{\cal F}_i^p(x_1,\mu) {\cal F}_j^{{\overline p}}(x_2,\mu)
\nonumber
\\ && ~~~~
\cdot {\hat \sigma}^{ij}_{NLO}(x_1,x_2,\mu)\,\,\,,
\end{eqnarray}
where $ {\cal F}_i^{p, {\overline p}}$ are the NLO parton distribution
functions for parton $i$ in a proton/antiproton, defined at a generic
factorization scale $\mu_f\!=\!\mu$, and ${\hat \sigma}^{ij}_{NLO}$ is
the ${\cal O}(\alpha_s^3)$ parton level total cross section for
incoming partons $i$ and $j$, made of the two channels $q\bar q,\,
gg\rightarrow t\bar t h$, and renormalized at an arbitrary scale
$\mu_r$ which we also take to be $\mu_r\!=\!\mu$.  At the Tevatron,
for $p {\overline p}$ collisions at hadronic center of mass energy
$\sqrt{s_H}\!=\!2$~TeV, more than $90\%$ of the tree level total cross
section comes from $q {\overline q}\rightarrow t {\overline t}h$,
summed over all light quark flavors.  Therefore, we compute $\sigma(p
{\overline p}\rightarrow t {\overline t}h)_{NLO}$ by including in
${\hat \sigma}^{ij}_{NLO}$ only the ${\cal O}(\alpha_s)$ corrections
to $ q {\overline q} \rightarrow t {\overline t} h$. The calculation
of $g g \rightarrow t {\overline t} h$ at ${\cal O}(\alpha_s^3)$ is,
however, crucial to determine $\sigma_{NLO}(pp\rightarrow t {\overline
  t} h)$ for the LHC, since in $pp$ collisions at
$\sqrt{s_H}\!=\!14$~TeV a large fraction of the total cross section
comes from the $gg\rightarrow t {\overline t}h$ channel.  The ${\cal
  O}(\alpha_s^3)$ total cross section for the LHC has
been estimated within the Effective Higgs Approximation in
Ref.~\cite{Dawson:1998im}. Full results are
presented in Ref.~\cite{been_etal} and will also appear in
Ref.~\cite{ggtth}.

We write the ${\cal O}(\alpha_s^3)$ parton level total cross section
as:
\begin{eqnarray}
&&{\hat \sigma}_{NLO}^{ij}(
x_1,x_2,\mu)=\\
&&=\alpha_s^2(\mu)
\biggl\{{\hat f}_{LO}^{ij}(
x_1,x_2)+
{\alpha_s(\mu)\over 4 \pi}
{\hat f }_{NLO}^{ij}(
x_1,x_2,\mu)\biggr\}\nonumber \\
&&\equiv{\hat \sigma}_{LO}^{ij}(
x_1,x_2,\mu)+
\delta {\hat \sigma}_{NLO}^{ij}(
x_1,x_2,\mu)\,\,\,,\nonumber
\end{eqnarray}
where $\alpha_s(\mu)$ is the strong coupling constant renormalized at
the arbitrary scale $\mu_r\!=\!\mu$, ${\hat \sigma}_{LO}^{ij}(
x_1,x_2,\mu)$ is the ${\cal O}(\alpha_s^2)$ Born cross section, and
$\delta {\hat \sigma}_{NLO}^{ij}(x_1,x_2,\mu)$ consists of the ${\cal
  O}(\alpha_s)$ corrections to the Born cross section. The Born cross
section ${\hat \sigma}_{LO}^{ij}(x_1,x_2,\mu)$ has a strong
$\mu$-dependence, which is canceled at NLO by $\delta {\hat
  \sigma}_{NLO}^{ij}( x_1,x_2,\mu)$, up to term of ${\cal
  O}(\alpha_s^4)$.  The resulting NLO cross section is therefore much
more stable under variations of $\mu$, as will be discussed in the
following (see also Fig.~\ref{fg:mudep}).

$\delta {\hat \sigma}_{NLO}^{ij}( x_1,x_2,\mu)$ contains both virtual
and real corrections to the lowest order cross section and can be
written as the sum of two terms:
\begin{eqnarray}
\label{terms}
\delta {\hat \sigma}_{NLO}^{ij}(
x_1,x_2,\mu)&= &\int d(PS_3) M(i j \rightarrow
t {\overline t} h)\nonumber \\
&+&
\int d(PS_4)M(i j \rightarrow
t {\overline t} h+g)\nonumber\\
&=& \sigma_{virt}+\sigma_{real}\,\,\,,
\end{eqnarray}
where $M(ij \rightarrow t\bar th )$ and $M(ij \rightarrow t\bar th+g)$
are respectively the matrix elements squared for the ${\cal
  O}(\alpha_s^3)$ $2\rightarrow 3$ and $2\rightarrow 4$ processes
averaged over the initial degrees of freedom and summed over the final
ones, while $d(PS_3)$ and $d(PS_4)$ denote the integration over the
corresponding three/four particle phase space.  The first term
represents the contribution of the virtual corrections, while the last
one is due to the real gluon emission.  \vspace{0.5truecm}

%\section{Virtual Corrections}
{\bf 3.} The ${\cal O}(\alpha_s)$ virtual corrections to the tree
level $ q {\overline q}\rightarrow t {\overline t} h$ process consist
of self-energy, vertex, box, and pentagon diagrams.  The calculation
of the virtual diagrams has been performed using dimensional
regularization in $d\!=\!4-2\epsilon$ dimensions. The diagrams have
been evaluated using FORM \cite{form} and \emph{Maple}, and all tensor
integrals have been reduced to linear combinations of a fundamental
set of scalar integrals.  We have computed analytically all scalar
integrals which give rise to either ultraviolet or infrared
singularities, while finite scalar integrals have been evaluated using
standard packages~\cite{vanOldenborgh:1990wn}.  Among the integrals
which have been computed analytically, many scalar box integrals and
the scalar pentagon integrals are extremely laborious, due to the
large number of massive particles present in the final state.  Box and
pentagon diagrams are ultraviolet finite, but have infrared
divergences.  These integrals are evaluated using the method of
Ref.~\cite{Bern:1993em} and analytic results are presented in
Ref.~\cite{lr2}.

Self-energy and vertex diagrams contain both infrared and ultraviolet
divergences. The ultraviolet divergences are renormalized by
introducing a suitable set of counterterms.  Since the cross section
is a renormalization group invariant, we only need to renormalize the
wave function of the external fields, the top quark mass, and the
coupling constants. We use on-shell subtraction for the wave-function
renormalization of the external fields. We define the top mass
counterterm in such a way that $m_t$ is the pole mass.  This
counterterm must be used twice: once to renormalize the top quark
mass, and again to renormalize the top quark Yukawa coupling.
Finally, for $\alpha_s(\mu)$ we use the $\overline{MS}$ scheme,
modified to decouple the top quark
\cite{Collins:1978wz}.  The first $n_{lf}$ light
fermions are subtracted using the $\overline{MS}$ scheme, while the
divergences associated with the heavy quark loop are subtracted at
zero momentum.\vspace{0.5truecm}

%\section{Real Corrections}

{\bf 4.} The ${\cal O}(\alpha_s)$ corrections to the Born cross
section due to real gluon emission have been computed using a two
cut-off implementation of the phase space slicing
algorithm~\cite{Harris:2001sx}. The contributions to $q {\overline q}
\rightarrow t {\overline t} h +g$ are first divided into a soft and a
hard contribution,
\begin{equation}
\label{eq:soft_hard}
\sigma_{real}(q {\overline q} \rightarrow t {\overline t} h+g)=
\sigma_{soft}+\sigma_{hard}\,\,\,,
\end{equation}
where \emph{soft} and \emph{hard} refer to the energy of the radiated
gluon. This division into hard and soft contributions depends on an
arbitrary soft cut-off, $\delta_s$, such that the energy of the
radiated gluon is considered soft if $E_g \le\delta_s {\sqrt{s}\over
  2}$. The cut-off $\delta_s$ must be very small, such that terms of
order $\delta_s$ can be neglected.  Therefore, to evaluate the soft
contribution, the eikonal approximation to the matrix elements can be
taken and the integral over the soft degrees of freedom performed
analytically.

The hard contribution to $q {\overline q} \rightarrow t {\overline t}
h+g$ is further divided into a hard/collinear and a hard/not collinear
region. The hard/collinear region is defined as the region where the
energy of the gluon is $E_g>\delta_s {\sqrt{s}\over 2}$ and the gluon
is radiated from the initial massless quarks at an angle $\theta_{ig}$
($i\!=\!q,\bar q$) such that $(1-\cos\theta_{gi})\le\delta_c$, for an
arbitrary small collinear cut-off $\delta_c$.  The matrix element
squared in the hard/collinear limit is found using the leading pole
approximation and the integration over the angular degrees of freedom
is performed analytically.  The hard gluon emission from the final
massive quarks never belongs to the hard/collinear region.  The
contribution from the hard/not collinear region is finite and is
computed numerically.

$\sigma_{soft}$ and $\sigma_{hard/coll}$ contain IR singularities,
which are calculated using dimensional regularization, and cancel
exactly the analogous singularities from the virtual contributions,
after absorbing mass singularities in the renormalized parton
distribution functions.

Both $\sigma_{soft}$ and $\sigma_{hard}$ depend on the two arbitrary
cut-offs $\delta_s$ and $\delta_c$, but their sum, i.e. the physical
cross section, is cut-off independent.  In Figs.~\ref{fg:deltas}, we
show the dependence of $\sigma_{real}$ on the soft cut-off,
$\delta_s$, for a fixed value of the hard/collinear cut-off,
$\delta_c\!=\!10^{-4}$. In the upper window we illustrate the
cancellation of the $\delta_s$ dependence between
$\sigma_{soft}+\sigma_{hard/coll}$ and $\sigma_{hard/non-coll}$, while
in the lower window we show $\sigma_{real}$ with the statistical errors
from the Monte Carlo integration. For $\delta_s$ in the range
$10^{-4}-10^{-2}$, a clear plateau is reached and the result is
independent of $\delta_s$.  We point out that Fig.~\ref{fg:deltas}
only shows distributions of the form $d\sigma/d\delta_s$, so the
corresponding cross sections are in this case twice the values that
can be read from the plot.  Analogously, Fig.~\ref{fg:deltac} shows
the independence of $\sigma_{real}$ on the hard/collinear cut-off,
$\delta_c$. All the results presented in the following are obtained
using $\delta_s$ and $\delta_c$ of order $10^{-3}$.
\begin{figure}[htb]
\begin{center}
\vspace*{0.8cm}
\hspace*{-1.5cm}
\epsfxsize=7cm \epsfbox{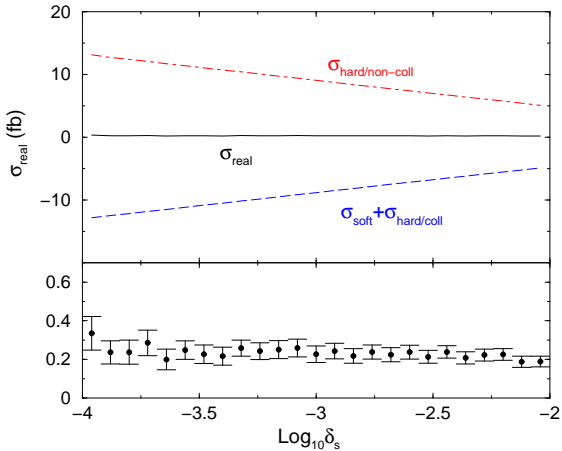}
\caption[ ]{ Dependence of $\sigma_{real}(p {\overline p}\rightarrow
  t {\overline t} h)$ on the soft cut-off $\delta_s$, at
  $\sqrt{s_H}\!=\!2$~TeV, for $M_h\!=\!120$ GeV, $\mu\!=\!m_t$, and
  $\delta_c=10^{-4}$. The lower scale shows the statistical error on
  $\sigma_{real}$.}
\label{fg:deltas}
\end{center}
\end{figure}

\begin{figure}[htb]
\begin{center}
\hspace*{-1.5cm}
%\vspace{.8cm}
\epsfxsize=7cm \epsfbox{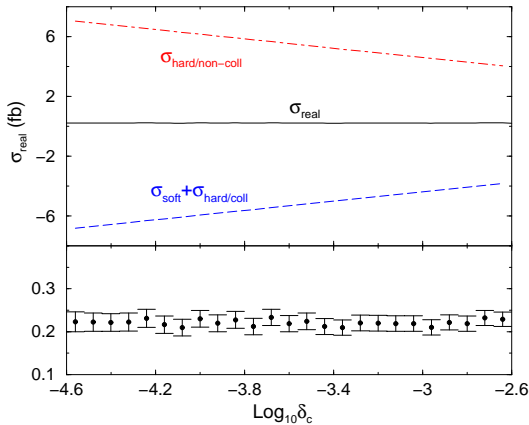}
\caption[ ]{Dependence of $\sigma_{real}(p {\overline p}\rightarrow
  t {\overline t} h)$ on the collinear cut-off $\delta_c$, at
  $\sqrt{s_H}\!=\!2$~TeV, for $M_h\!=\!120$ GeV, $\mu\!=\!m_t$, and
  $\delta_s\!=\!0.005$. The lower scale shows the statistical error on
  $\sigma_{real}$.}
\label{fg:deltac}
\end{center}
\end{figure}

%\section{Results}
{\bf 5.}
Our numerical results are found using CTEQ4M parton distribution
functions for the calculation of the NLO cross section, and CTEQ4L
parton distribution functions for the calculation of the lowest order
cross section ~\cite{Lai:1997mg}.  The NLO (LO) cross section is evaluated
using the $2$ ($1$)-loop evolution of $\alpha_s(\mu)$. The top quark
mass is taken to be $m_t\!=\!174$~GeV and
$\alpha_s^{NLO}(M_Z)\!=\!.116$.
 
First of all, in Fig.~\ref{fg:mudep} we show, for $M_h\!=\!120$~GeV,
how at NLO the dependence on the arbitrary renormalization scale $\mu$
is significantly reduced. We notice that only for scales $\mu$ greater
than $2 m_t$ is the NLO result greater than the lowest order result.
\vspace{0.5truecm}

\begin{figure}[hbt]
\begin{center}
\hspace*{-1.5cm}
\epsfxsize=7cm \epsfbox{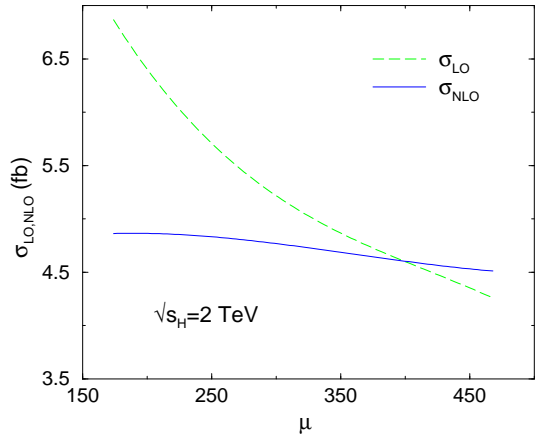}
\caption[ ]{ Dependence of $\sigma_{LO,NLO}(p {\overline p}\rightarrow
  t {\overline t} h)$ on the renormalization scale $\mu$, at
  $\sqrt{s_H}\!=\!2$~TeV, for $M_h\!=\!120$ GeV. }
\label{fg:mudep}
\end{center}
\end{figure}

Fig.~\ref{fg:signlo} shows both the LO and the NLO total cross section
for $p\bar p\rightarrow t\bar th$ at $\sqrt{s_H}\!=\!2$~TeV, for
two values of the renormalization scale $\mu\!=\!m_t$ and $\mu\!=\!2
m_t$.  Over the entire range of $M_h$ accessible at the Tevatron, the
NLO corrections decrease the rate.  For example, for $M_h\!=\!120$~GeV
and $\mu\!=\!m_t$ the NLO total cross section is reduced to $4.86\pm
0.03~fb$ from the lowest order prediction of $6.868\pm 0.002~fb$. The
reduction is much less dramatic at $\mu\!=\!2m_t$, as can be seen from
both Fig.~\ref{fg:mudep} and Fig.~\ref{fg:signlo}. The complete NLO
result includes the lowest order result and the contribution of the
two terms given in Eq.~(\ref{terms}). The error we quote on our values
is the statistical error on the numerical integration involved in
evaluating the total cross section.

The corresponding K-factor, i.e. the ratio of the NLO
cross section to the LO one
\begin{equation}
K=\frac{\sigma_{NLO}}{\sigma_{LO}}
\end{equation}
is shown in Fig.~\ref{fg:sigk}. Given the strong scale dependence of
the LO cross section, the K-factor also shows a significant
$\mu$-dependence, while it is almost constant with $M_h$. For scales
$\mu$ between $\mu\!=\!m_t$ and $\mu\!=\!2m_t$, the K-factor varies
roughly between $K\!=\!0.70$ and $K\!=\!0.95$.  The reduction of the
NLO cross section with respect to the Born cross section is due to
fact that at $\sqrt{s_H}\!=\!2$ TeV the $t\bar th$ final state is
produced in the threshold region. In this region the gluon exchange
between the final state quarks gives origin to Coulomb singularities
that contribute to the cross section with terms of order
$\alpha_s/\beta$, where $\beta$ is the velocity of the top/antitop
quark in the $t\bar t$ CM frame.  Since the $t\bar th$ final state is
in a color octet configuration, these corrections are negative and
therefore decreases the Born cross section, causing the K-factor to be
smaller than unity.  The same effect was observed in the NLO cross
section for $e^+e^-\rightarrow t\bar th$~\cite{Dittmaier:1998dz}. In
that case, however, the $t\bar th$ final state is in a color singlet
configuration and the threshold corrections are positive.
\begin{figure}[htb]
\begin{center}
\vspace*{.5cm}
\hspace*{-1.5cm}
\epsfxsize=7cm \epsfbox{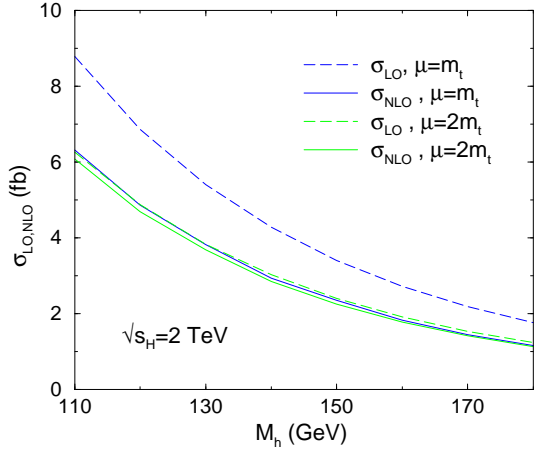}
\caption[ ]{$\sigma_{NLO}$ and $\sigma_{LO}$  for
  $p {\overline p} \rightarrow t {\overline t} h$ as functions of
  $M_h$, at $\sqrt{s_H}\!=\!2$~TeV, for $\mu\!=m_t$ and
  $\mu\!=\!2m_t$.}
\label{fg:signlo}
\end{center}
\end{figure}
\begin{figure}[htb]
\begin{center}
%\vspace*{0.8cm}
\hspace*{-1.5cm}
\epsfxsize=7cm \epsfbox{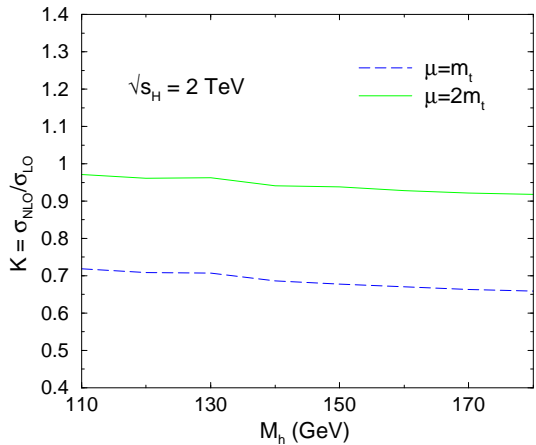}
\caption[ ]{K factor for $p
  {\overline p}\rightarrow t {\overline t} h$ as a function of $M_h$,
  at $\sqrt{s_H}\!=\!2$~TeV, for $\mu\!=m_t$ and $\mu\!=\!2m_t$ .}
\label{fg:sigk}
\end{center}
\end{figure}

%\section{Conclusion}
{\bf 6.}  The next-to-leading order QCD corrections to the Standard
Model process $p {\overline p}\rightarrow t {\overline t} h$, at
$\sqrt{s_H}\!=\!2$ TeV, reduce the cross section by a factor of
$0.7-0.95$ for renormalization and factorization scales $m_t
\!<\!\mu\!<\!2 m_t$. The NLO result shows a drastically reduced scale
dependence as compared to the Born result and leads to increased
confidence in predictions based on these results.

\section*{Acknowledgements}
We thank Z.~Bern, F.~Paige, and D.~Wackeroth for valuable discussions
and encouragement. We are grateful to the authors of
Ref.~\cite{been_etal} for detailed comparisons of results prior to
publication.  The work of L.R. (S.D.) is supported in part by the U.S.
Department of Energy under grant DE-FG02-97ER41022
(DE-AC02-76CH00016).
\end{narrowtext}
%\vspace{-0.7truecm}

\end{document}